\documentclass[showpacs,prl,aps,twocolumn]{revtex4-1}
\usepackage{graphicx}
\usepackage{amsmath}
\usepackage{amssymb}
\usepackage{amsthm}
\usepackage{bbm}
\usepackage{mathrsfs}
\usepackage{bm}

\newcommand{\Rc}{R}
\newcommand{\mESO}{\text{effective SO}}

\begin{document}
\title{Many-body out-of-equilibrium dynamics of hard-core lattice bosons with non-local loss}
\author{B. Everest}
\author{M. R. Hush}
\author{I. Lesanovsky}
\affiliation{School of Physics and Astronomy, University of Nottingham, Nottingham, NG7 2RD, United Kingdom}

\begin{abstract}
We explore the dynamics of hard-core lattice bosons in the presence of strong non-local particle loss. The evolution occurs on two distinct time-scales, first a rapid strongly correlated decay into a highly degenerate Zeno state subspace, followed by a slow almost coherent evolution. We analytically solve the fast initial dynamics of the system, where we specifically focus on an initial Mott insulator state, and perform an analysis of the particle arrangements in the Zeno subspace. We investigate the secondary slow relaxation process that follows and find an intricate regime where the competition between dissipation and coherence results in various types of interacting particle complexes. We classify them and analyse their spectral properties in the presence and absence of nearest-neighbor interactions. Under certain circumstances the dispersion relations of the complexes feature flat bands, which are a result of an effective spin-orbit coupling.
\end{abstract}
\maketitle

\textit{Introduction.---}The out-of-equilibrium behaviour of open quantum many-body systems is currently under intense investigation \citep{Ates2012,Diehl2010a,Kollath2007,Lemeshko2013,Genway2014}. This interest is rooted in the fact that often the competition between coherent and incoherent processes gives rise to seemingly counterintuitive phenomena. Examples are the creation of entanglement by dissipation \cite{Kraus2008,Maniscalco2008,Verstraete2009,Zhu2014,Brazhnyi2009,Weimer2010,Lin2013,Caballar2014,Orszag2001} and the emergence of effective interparticle interactions \cite{Syassen2008,Garcia-Ripoll2009}. In certain cases the latter may even lead to a binding mechanism \cite{Ates2012,Lemeshko2013}, which is qualitatively different to the one resulting from coherent forces that bind constituent particles, in for example molecules or atoms \cite{DeKock1989}. In Ref. \cite{Ates2012} the creation of dissipatively bound complexes was shown to be due to the quantum Zeno effect \cite{Misra1977,Maniscalco2008,Itano1990,Pascazio1994,Frerichs1991,Facchi2004}, i.e. due to strong dissipation preventing the occupation of particular states by projecting the system onto a reduced state space, the Zeno subspace. While this leads to a good understanding of the few-body physics, a systematic exploration of out-of-equilibrium dynamics on the many-body level is so far lacking.

The purpose of this work is to provide insight into many-body dynamics resulting from a competition between coherent particle motion and strong non-local particle loss through primarily analytic analysis. To this end we consider the situation of a one-dimensional lattice filled with hard-core bosons in a Mott insulating state. We find that the evolution proceeds in two stages. The first stage is characterized by a purely dissipative dynamics that leads to a strongly correlated loss of bosons until the system reaches a highly degenerate Zeno subspace. The second stage is governed by the competition between the dissipation and coherent particle hopping that leads to the formation of dissipatively bound complexes. We identify two qualitatively different types which naturally occur in the Zeno subspace. Their dispersion relations depend strongly on the number of constituent bosons and we find for some configurations the emergence of so-called flat bands \cite{Wu2007} which result from an effective spin-orbit coupling and gives rise to immobile complexes \cite{Takahashi2013}. Such flat bands are of interest in the study of exotic topological states of matter e.g. in quantum Hall physics \cite{Wang2011}. We further analyze the effect of interactions among neighboring bosons and between complexes.

\begin{figure}
\begin{center}
\includegraphics[scale=0.19]{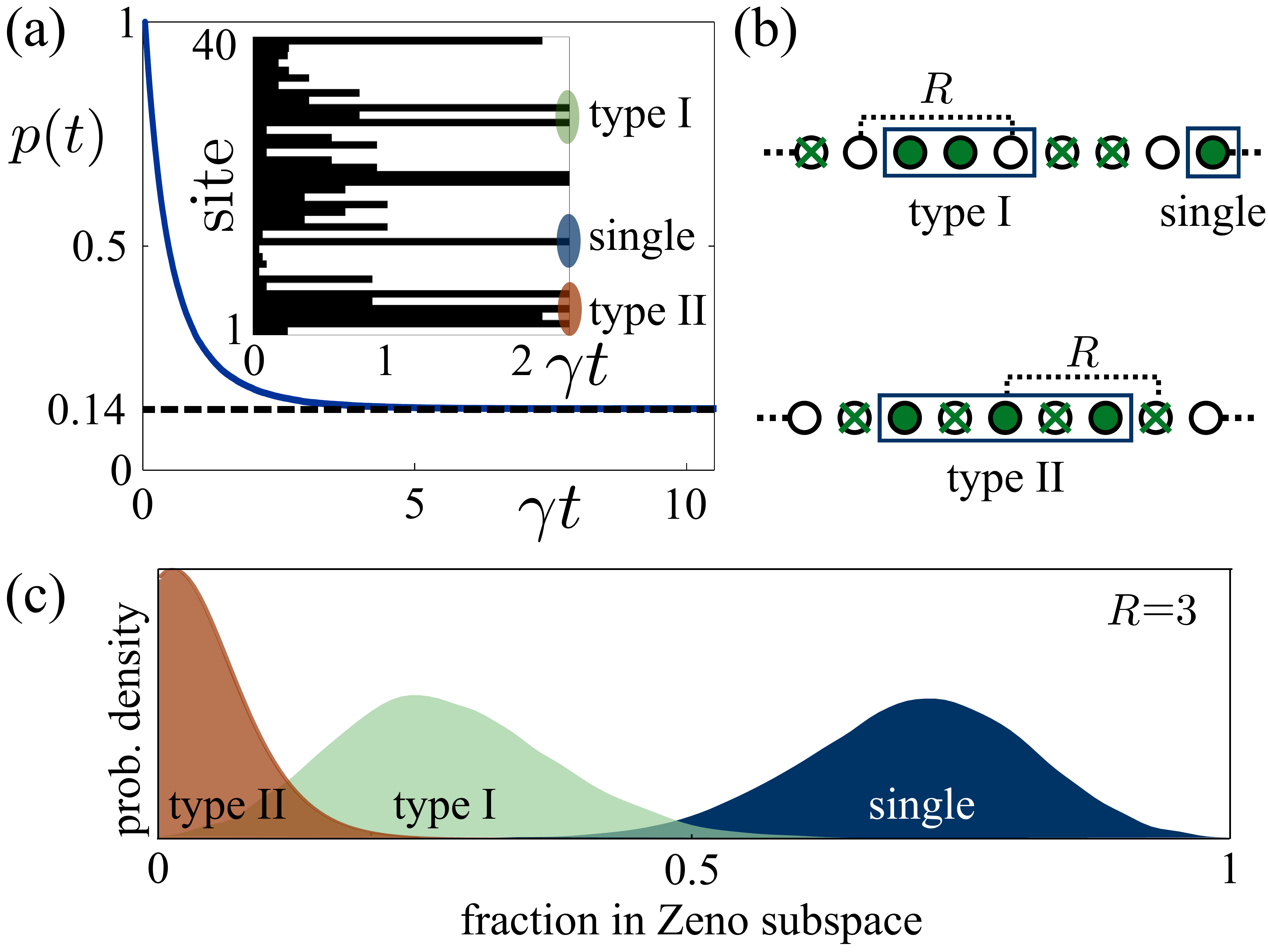}
\caption{All figures are for $\Rc=3$. (a) Evolution of the boson density $p(t)$ under the dissipative dynamics $\mathcal{L}_d$ from an initial Mott insulator. The stationary density is $p(t\rightarrow\infty)=e^{-2}\approx 0.14$. The inset shows a single trajectory with 40 bosons and periodic boundary conditions. For this particular trajectory 16 jumps (loss processes) occur, such that the final state contains solely 8 bosons. (b) Representative boson arrangements in the stationary state, where single free bosons and two types of particle complexes can emerge. The circles indicate sites, a filled circle indicates an occupied site, a cross indicates a site whose occupation is forbidden, as the resulting configuration would not lie in the Zeno subspace, and a box indicates the ``size" of a complex. The type I complex --- defined as having a size smaller than $\Rc$ --- is, in this example, constituted of two bosons. These bosons are unable to tunnel away from each other without running into a forbidden site which leads to an effective binding. The type II complex has a spatial extent that is larger than $\Rc$. It is qualitatively different to type I in the sense that the removal of one boson (in the center) destroys the binding for the remaining ones. (c) Probability distributions for single bosons, type I and type II complexes in the stationary state that is reached from a Mott insulator.}
\label{fig:initialCondition}
\end{center}
\end{figure}

\textit{System.---}We consider a one-dimensional lattice with $N$ sites filled with hard-core bosons \cite{Garcia-Ripoll2009}, a scenario which can for example be realized with optically trapped cold atoms \cite{Bloch2008}. Bosons tunnel between adjacent sites at a rate $J$ such that the Hamiltonian is given by $H=J\sum_j (\sigma^-_j\sigma^+_{j+1} + \sigma^+_j\sigma^-_{j+1})$. Here $\sigma^\pm_k=(\sigma^x_k \pm i\sigma^y_k)/2$, with $\{\sigma^x,\sigma^y,\sigma^z\}$ being the standard Pauli matrices. In addition to the Hamiltonian evolution we consider non-local dissipation which is given by distance-selective pair loss, meaning that two bosons separated by the critical distance $\Rc$ are ejected from the lattice at a rate $\gamma$ [see Fig. \ref{fig:initialCondition}]. This type of dissipation can be physically realised in cold atoms experiments by exploiting the properties of high-lying excited states, so-called Rydberg states, as shown in Ref. \cite{Ates2012}. The dynamics of the density matrix $\rho$ of the system is described by a master equation in Lindblad form, $\dot{\rho} = -i[H,\rho] + \sum_{j=1}^N (L_j\rho L_j^\dagger - \frac{1}{2}\lbrace L_j^\dagger L_j , \rho\rbrace)\equiv\mathcal{L}_c\,\rho+ \mathcal{L}_d\,\rho$, with jump operators $L_j = \sqrt{\gamma}\sigma^-_j \sigma^-_{j+\Rc}$. In this work we focus on the limit of strong dissipation, e.g. $\gamma \gg J$. This leads to a separation of the two timescales on which the coherent $\mathcal{L}_c$ and dissipative $\mathcal{L}_d$ dynamics proceed.

\textit{Fast dissipative dynamics and the Zeno subspace.---}We begin by analyzing the fast dissipative dynamics. Its stationary subspace --- the Zeno subspace --- is spanned by all states $\left|s\right>$ that satisfy $L_j\left|s\right>=0\, \forall\, j$, i.e. they do not contain any two bosons at the critical distance $\Rc$. To understand the dissipative non-equilibrium evolution into the Zeno subspace we consider our system starting in a Mott insulator state. The corresponding evolution of the average boson density $p(t)=\sum_j \left<n_j\right>(t)/N$, with $n_j=\sigma^+_j\sigma^-_j$, can be found analytically: The mean value of the density on site $j$ evolves under the fast dynamics of $\mathcal{L}_d$ according to $\dot{\langle n_j \rangle} = -\gamma(\langle n_j n_{j+\Rc} \rangle+ \langle n_{j-\Rc} n_j \rangle)$, i.e. it depends on a two-point correlation function. Defining the correlators $C_k = \langle\prod_{l=0} ^k n_{j+l\Rc}\rangle $ and using translational invariance we obtain the hierarchy $\dot{C}_k =  -\gamma (kC_k + 2C_{k+1})$. This equation can be solved by introducing (see Ref. \cite{Gardiner}) the generating function $G(x,t) = \sum_{k=0}^{\infty} x^kC_k/k! $ which evolves according to $\dot{G}(x,t) = -\gamma (2+x)\partial_x G(x,t)$.

For a Mott insulator state we have the initial condition $C_k=1$ and therefore $G(x,0)=\sum_{k=0}^{\infty} x^k/k!=e^x$. With this, the general solution becomes $G(x,t) = e^{(2+x)e^{-\gamma t}-2}$ and thus the density evolves as $p(t)  = C_0= G(x=0,t) = e^{2(e^{-\gamma t}-1)}$. Numerical Monte Carlo simulations [see Fig. \ref{fig:initialCondition}(a)] confirm the rapid exponential decay of the boson density on a timescale $\sim \gamma^{-1}$. The inset shows a generic trajectory which displays the fast removal of boson pairs and a stationary configuration in which boson pairs at distance $\Rc$ are absent. This is one configuration of many that span the high dimensional stationary Zeno subspace, the projector onto which can be explicitly written as $Q_0=\prod_{j=1}^N (1-n_jn_{j+\Rc})$. The average density in the stationary state reached from a Mott insulator is given by $p(t\rightarrow\infty) = e^{-2} \approx 0.14$. Note, that this calculation is in fact \emph{independent} of the value of $\Rc$

\textit{Effective coherent dynamics in the Zeno subspace.---}Once having reached the Zeno subspace the dissipative evolution governed solely by $\mathcal{L}_d$ comes to a halt. However, in the presence of quantum tunneling, due to $\mathcal{L}_c$, non-trivial coherent dynamics emerges which takes place on a timescale $J^{-1}$. As shown in Ref. \cite{Ates2012} the effective master equation for the projected density matrix onto the Zeno subspace, $\mu\equiv Q_0\rho Q_0$, in the limit $\gamma\gg J$, is obtained by means of Kato perturbation theory \cite{Garcia-Ripoll2009,Kato}: $\dot{\mu}=-i[H_\text{Z},\mu]+ \sum_{j,\alpha} (L^{(Z)}_{j,\alpha}\mu {L^{(Z)}_{j,\alpha}}^\dagger - \frac{1}{2}\lbrace {L^{(Z)}_{j,\alpha}}^\dagger {L^{(Z)}_{j,\alpha}} , \mu\rbrace)$, with
\begin{eqnarray*}
H_\text{Z} &= & Q_0 H Q_0, \label{eqn:effectiveHamiltonian} \\
L_{j,1}^{(Z)} &= & \sqrt{2\Gamma}(A_j - \sigma_{j-{\Rc}}^{+}B_j-\sigma_{j+2{\Rc}}^{+}B_{j+{\Rc}}) \label{eqn:effectiveL1} \\
L_{j,2}^{(Z)} &= & \sqrt{\Gamma}B_j, \label{eqn:effectiveL2}
\end{eqnarray*}
with the effective decay rate $\Gamma=2J^2/\gamma$ and the operators $A_j = \sigma_{j+\Rc+1}^- \sigma_{j}^- + \sigma_{j+\Rc-1}^- \sigma_{j}^- + \sigma_{j+\Rc}^- \sigma_{j+1}^- + \sigma_{j+\Rc}^- \sigma_{j-1}^-$ and $B_j = \sigma_{j-\Rc}^- \sigma_{j-1}^- \sigma_{j+\Rc}^- + \sigma_{j-\Rc}^- \sigma_{j+1}^- \sigma_{j+\Rc}^-$.

By construction the dynamics under $H_\text{Z}$ is restricted to the Zeno subspace. Dissipation within the Zeno subspace affects boson pairs ($L_{j,1}^{(Z)}$) or triples ($L_{j,2}^{(Z)}$) in configurations that are ``one tunneling event away" from containing bosons at the critical distance $\Rc$. Such configurations undergo an incoherent evolution at a rate $\Gamma$, which is strongly suppressed for fast two-body decay $\gamma\gg J$. Therefore the evolution within the Zeno subspace becomes predominantly coherent.

\textit{Families of coherent particle complexes.---}The approximately coherent evolution under $H_\text{Z}$ has interesting consequences. Due to the explicit appearance of the projector $Q_0$, the simultaneous occupation of two sites at a distance of $\Rc$ is forbidden. This leads to strong correlations and the formation of bound complexes. These complexes can contain a variable number of bosons, but there are two qualitatively different configuration sets in which they can form. Let us start with the simplest case --- referred to as \emph{type I} --- aspects of which were already discussed in Ref. \cite{Ates2012}. Here $m$ bosons are localized in a region with spatial extent smaller than $\Rc$, an example of which is shown in Fig. \ref{fig:initialCondition}(b). These bosons are effectively bound since they cannot separate by more than $\Rc - 1$ sites under the evolution governed by $H_\text{Z}$. The second class --- \emph{type II} --- are distinguished by having a spatial extent greater than $\Rc$. These complexes can form when the bosons and their associated critical distances overlap [see Fig. \ref{fig:initialCondition}(b)]. Here, unlike for type I, not every particle binds all the others, but one can even encounter situations in which the removal of one boson destroys the entire complex, an example of which is shown in Fig. \ref{fig:initialCondition}(b). Both type I and II complexes appear naturally in the stationary state that is reached from a Mott insulator. Their relative abundance is shown in the histogram presented in Fig. \ref{fig:initialCondition}(c). Besides single bosons, there is a significant proportion that occupy a type I state and only a small number enter a type II state. In the following we will perform a detailed investigation of their properties.

\textit{Type I complexes.---}We limit our study to the dynamics of a single complex in the lattice, addressing the interactions among complexes later. In the following we will provide three qualitatively different examples: immobile complexes without internal structure, complexes with an internal structure and effective spin-orbit (SO) coupling, and complexes whose dispersion relations feature a flat band arising from this $\mESO$ coupling.

We start with the simplest type I state: two bosons and a critical distance $\Rc=2$. The only possible configuration of these bosons, in a type I state, is to be adjacent. Thus, the basis states are $|j,1\rangle = \sigma_j^{+}\sigma_{j+1}^{+}|\Phi\rangle$, where $|\Phi\rangle$ is the vacuum state. In this notation $j$ denotes the position of the complex and the second index labels the ``internal state" of the complex. The projected Hamiltonian $H_\text{Z}$ in this basis is identically zero. Hence the basis states are trivially eigenstates and $|j,1\rangle$ represents immobile type I complexes. These type I solutions emerge whenever $\Rc = m$.

\begin{figure}[t!]
\begin{center}
\includegraphics[scale=0.19]{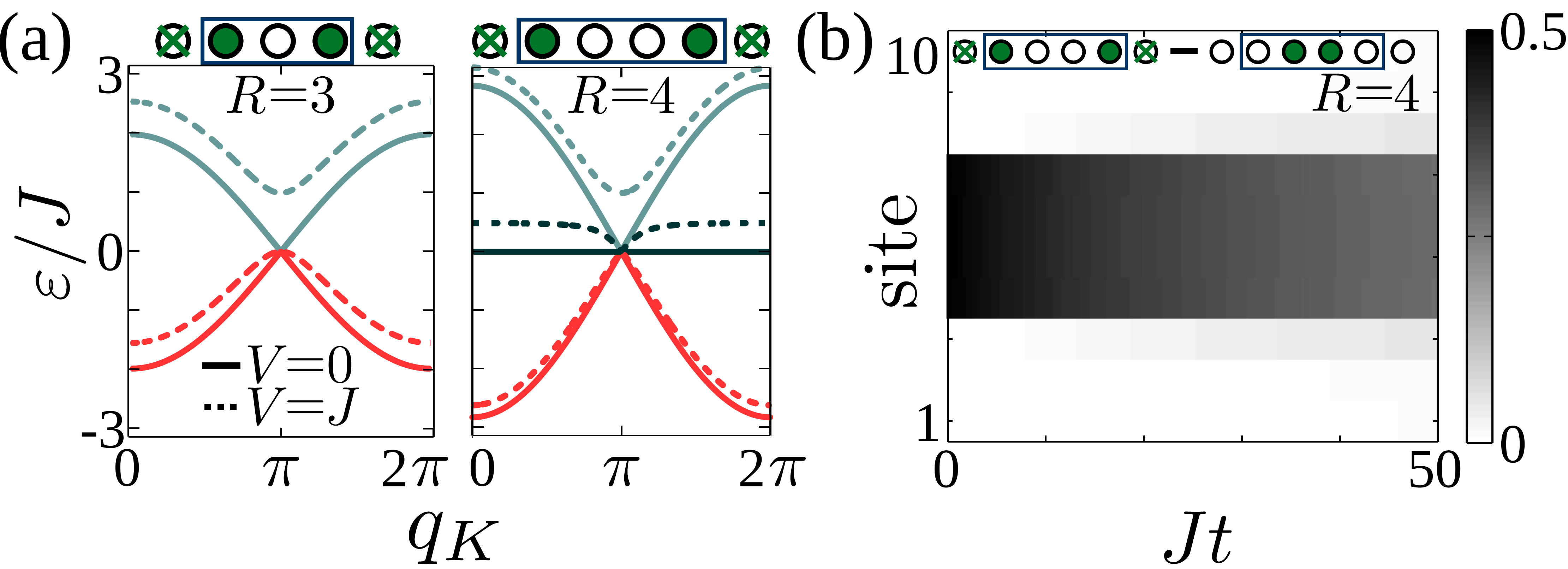}
\caption{(a) Dispersion relations (solid curves) for type I complexes of two bosons with $\Rc=3$ and $\Rc=4$. Both cases show a crossing at $q_K=\pi$, and when $\Rc=4$ a flat band occurs. In the presence of nearest neighbor interactions (here $V=J$) the degeneracy is lifted and the flat bands are distorted (dashed curves). The sketches above the panels show a particular internal state of the respective complex. Panel (b) shows the evolution of the boson density of a type I complex formed by two bosons in the state $|\text{F}^{(\text{I})}_j\rangle$ with $\Rc=4$ (see sketch above the panel) and $\gamma=100J$ on a lattice of $10$ sites simulated with the full master equation.}
\label{fig:dispersionAnd2Locked}
\end{center}
\end{figure}

In order to see some non-trivial physics we require a complex with some ``internal states". The simplest case of this is constituted by 2 bosons with $\Rc=3$, previously discussed in \cite{Ates2012}. In order to calculate the spectrum of this complex, a basis of the internal states is defined as $|j,1 \rangle = \sigma_j^+\sigma_{j+1}^+|\Phi\rangle$ and $|j, 2\rangle = \sigma_j^+\sigma_{j+2}^+|\Phi\rangle$. We may also define a creation operator $|j,\alpha \rangle \equiv b_j^{(\alpha)\dagger} |\Phi\rangle$, allowing us to express $H_\text{Z} = J\sum_j[ b_j^{(2)\dagger} b_j^{(1)} + b_{j+1}^{(1)\dagger} b_{j}^{(2)}  + \text{H.c.}]$. To obtain the corresponding dispersion relation $\varepsilon_\pm(K)$ [see Fig. \ref{fig:dispersionAnd2Locked}(a)] and eigenstates $|K_\pm\rangle$, we perform a discrete Fourier transform, using periodic boundary conditions and find: $\varepsilon_\pm(K) =  \pm 2J \cos\left(\frac{q_K}{2}\right)$, $|K\rangle_\pm =  \frac{1}{\sqrt{2N}} \sum_j e^{ijq_K}[|j,2 \rangle \pm e^{-iq_K /2}|j,1 \rangle]$, where $q_K = 2\pi K/N$ is the quasi-momentum. We see that the internal state of the complex is strongly linked to its motion on the lattice, namely the group velocity of the internal states is always in the opposite direction for the same quasi-momentum. This is what we term as $\mESO$ coupling. Note that this spectrum has a degeneracy or crossing that occurs at $q_K=\pi$.

Lastly we consider a complex where the $\mESO$ coupling results in a flat band, namely the case of two bosons with $\Rc = 4$. We define a basis with three internal states as: $|j,1\rangle = \sigma_j^+\sigma_{j+1}^+|\Phi\rangle$, $|j,2\rangle = \sigma_j^+\sigma_{j+2}^+|\Phi\rangle$ and $|j,3\rangle = \sigma_j^+\sigma_{j+3}^+|\Phi\rangle$. The resulting dispersion relations [shown in Fig. \ref{fig:dispersionAnd2Locked}(a)] and eigenstates are given by
\begin{eqnarray*}
\varepsilon_\eta(K) &= & \eta \,2\sqrt{2}J \cos\left(\frac{q_K}{2}\right), \label{eqn:dispersion24} \\
|K_0\rangle &= &  \sum_j \frac{e^{ijq_K}}{\sqrt{2N}} [|j,3\rangle - e^{-iq_K}|j,1\rangle], \label{eqn:1eigenstate24} \\
|K_\pm\rangle &= &\sum_j  \frac{e^{ijq_K}}{2\sqrt{N}} [e^{-iq_K}|j,1\rangle \pm \sqrt{2}e^{-iq_K /2}|j,2\rangle + |j,3\rangle]. \label{eqn:2eigenstate24}
\end{eqnarray*}
This complex has three branches labelled by $\eta=\{0,+,-\}$. The branch $\eta=0$ is a flat band. Dispersion relations featuring flat bands may result in immobile localized states which in contrast to the first type I example are non-trivial. Localized states are formed by superimposing many quasi-momentum eigenstates and hence for non-flat dispersion relations, immobile states cannot form. However, in a flat band all quasi-momentum states have the \emph{same} energy and the resulting superposition state is thus an eigenstate of the Hamiltonian.

A concrete example is given by the states $|\text{F}^{(\text{I})}_j\rangle=(\sqrt{2/\Rc})\sum_{i=j}^{j+\Rc/2-1}[(-1)^{i}\sigma_i^+ \sigma_{\Rc-i+1}^+]|\Phi\rangle$. Using one of these states as the initial condition and propagating it under the \emph{full master equation} we find indeed that it remains immobile as shown in Fig. \ref{fig:dispersionAnd2Locked}(b). Note, that the boson density is slowly decreasing on a timescale $\Gamma^{-1}$. This clearly shows that the flat bands are not an artifact of infinitely strong dissipation but instead that they indeed have a drastic effect on the boson dynamics in a system with competing coherent and dissipative evolution.

Let us make some general remarks on the emergence of flat bands in case of type I complexes: For complexes consisting of two bosons, flat bands exist provided that $\Rc$ is even. Furthermore, we find that for two, three and four bosons a flat band emerges when $\Rc/m\in \mathbb{N}$. Interactions among bosons also play an important role. In order to illustrate this we consider nearest-neighbor interactions of the form $H_{\text{nn}}=V\sum_j n_j n_{j+1}$ which might, for instance, emerge in cases where non-local loss is engineered via Rydberg dressing (see Ref. \cite{Ates2012}). Such interactions modify the dispersion relations as shown in Fig. \ref{fig:dispersionAnd2Locked}(a) in the sense that they lift the degeneracy point observed for $\Rc=3$, and distort the flat band in the case of $\Rc=4$.

\begin{figure}[t!]
\begin{center}
\includegraphics[scale=0.19]{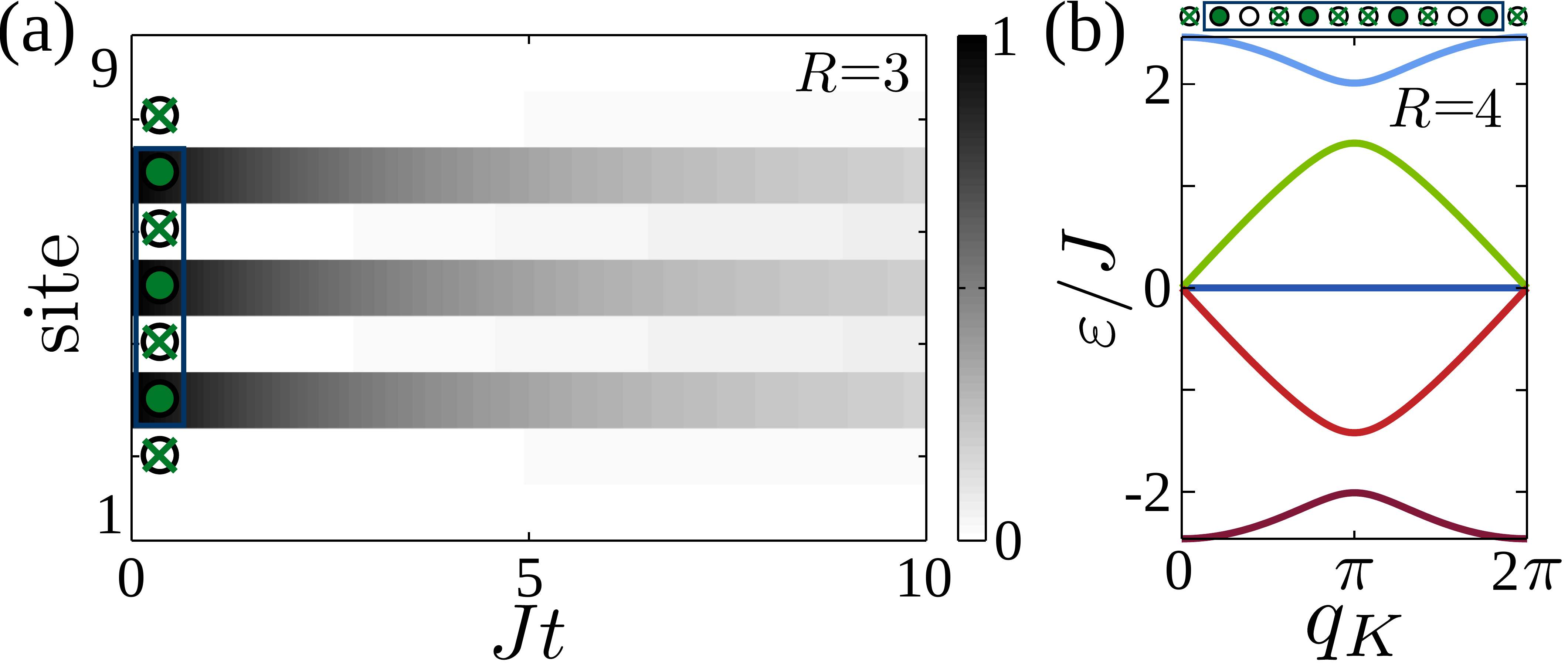}
\caption{(a) Evolution of boson density for a type II complex in the immobile state $|\text{F}^{(\text{II})}_3\rangle$, with $\Rc=3$ and $\gamma=100J$. (b) Dispersion relation for a type II complex consisting of four bosons with $\Rc=4$.}
\label{fig:lockedHardCore3AtomsX3}
\end{center}
\end{figure}
\textit{Type II complexes.---}We now move our study to type II complexes, i.e. complexes that are larger than $\Rc$. We give two examples, one without internal structure and one with $\mESO$ coupling.

First we consider three bosons and a critical distance $\Rc=3$. The only possible type II complexes have the basis $|j,1\rangle = \sigma_j^+\sigma_{j+2}^+\sigma_{j+4}^+|\Phi\rangle$. They are immobile --- similar to the first type I example --- as each boson's movement is inhibited by its the nearest bosons. This is confirmed as well by numerical exact simulations as shown in Fig. \ref{fig:lockedHardCore3AtomsX3}(a). Such immobile states can be straight-forwardly generalized to larger boson numbers, e.g. in the given example by attaching bosons to either end of the complex keeping a separation of one site.

In the second example we consider four bosons and a critical distance $\Rc = 4$. The resulting complex has five internal states: $|j,1\rangle = \sigma_j^+\sigma_{j+3}^+\sigma_{j+6}^+\sigma_{j+9}^+|\Phi\rangle$, $|j,2\rangle = \sigma_j^+\sigma_{j+3}^+\sigma_{j+6}^+\sigma_{j+8}^+|\Phi\rangle$, $|j,3\rangle = \sigma_j^+\sigma_{j+3}^+\sigma_{j+5}^+\sigma_{j+8}^+|\Phi\rangle$, $|j,4\rangle = \sigma_j^+\sigma_{j+2}^+\sigma_{j+5}^+\sigma_{j+7}^+|\Phi\rangle$ and $|j,5\rangle = \sigma_j^+\sigma_{j+2}^+\sigma_{j+5}^+\sigma_{j+8}^+|\Phi\rangle$ and the dispersion relations shown in Fig. \ref{fig:lockedHardCore3AtomsX3}(b): One is given by $\varepsilon_0(K)=0$ and the other four are $\varepsilon_{\eta,\delta}(K)=\eta\,\sqrt{3+\delta\sqrt{5+4\cos(q_K)}}$, with $\eta,\delta=\pm$. Hence, this type II complex features a flat band and spatially localized states of the form $|\text{F}^{(\text{II})}_j\rangle=[-\sigma_j^+\sigma_{j+3}^+\sigma_{j+6}^+\sigma_{j+9}^+ + \sigma_{j+1}^+\sigma_{j+3}^+\sigma_{j+6}^+\sigma_{j+8}^+]\Phi\rangle$.

Again let us conclude with some more general remarks: A flat band of similar structure exists for five bosons with $\Rc=4$. For $\Rc=3$ and $4$, a flat band exists provided the number of bosons is equal to or greater than $\Rc$. The dispersion relation of this type II complex is not modified by the presence of nearest neighbor interactions. This is due to the fact that given the arrangement of the bosons, the simultaneous occupation of neighboring sites is forbidden. Thus, the flat bands of certain type II complexes are in this case protected from interaction effects in contrast to the type I case.

\begin{figure}[t!]
\begin{center}
\includegraphics[scale=0.19]{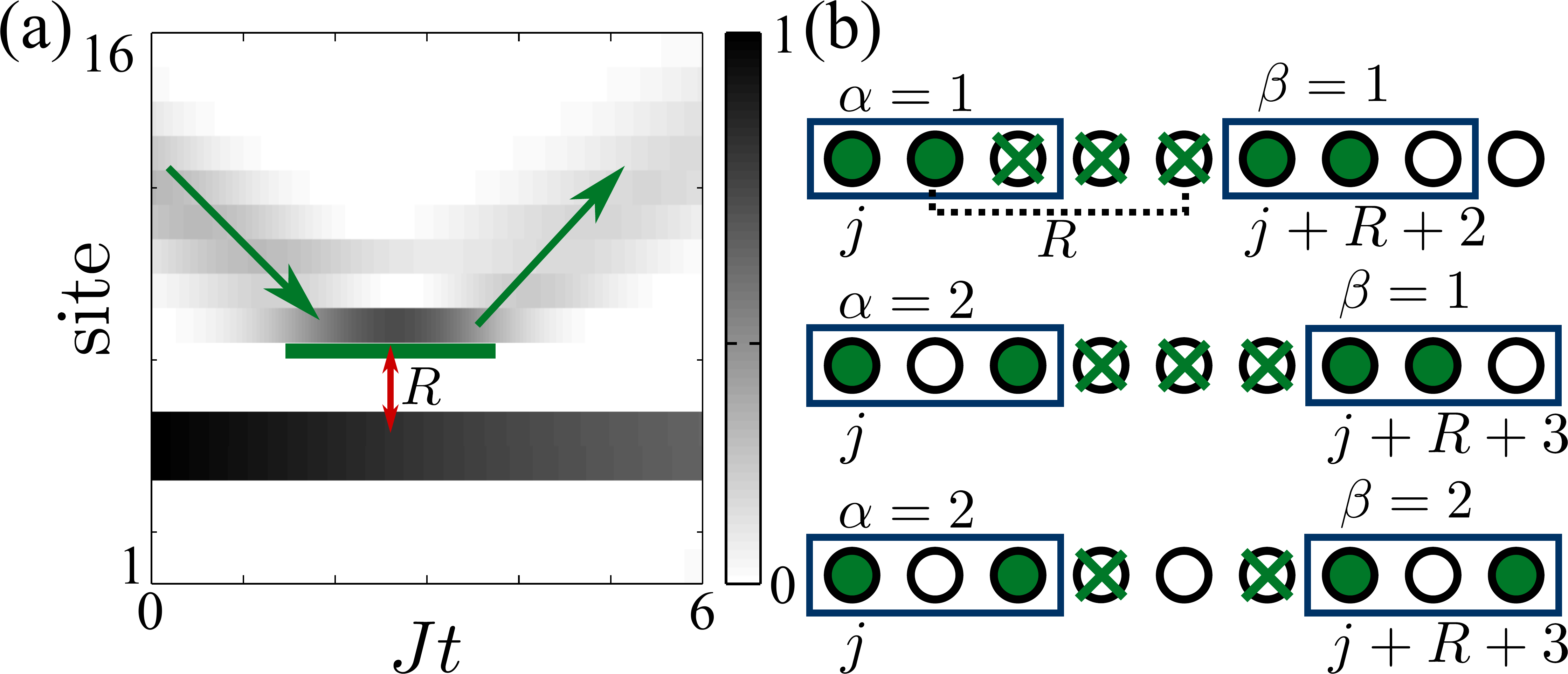}
\caption{(a) Evolution of the boson density for a single boson impinging an immobile type II complex ($\Rc=2$). The single boson is in the wave packet state $|\text{G}\rangle$ with initial central quasi-momentum of $q_0=\pi/2$ and width $\sigma=2$. The two-particle loss rate is $\gamma=100J$. The single boson is reflected elastically off the type II complex due to the presence of an effective next nearest neighbor exclusion interaction. (b) We show three examples of two type I complexes, in different internal states, interacting with one another. We see that the distance of the interaction depends on the internal state of the complexes.}
\label{fig:bouncingStates}
\end{center}
\end{figure}

\textit{Interaction between complexes.---}As can be seen in the inset of Fig. \ref{fig:initialCondition}(a) complexes are typically not isolated in the stationary subspace of $\mathcal{L}_d$. Hence, interactions between complexes, and complexes and single bosons occur. Given the abundance of each species [see Fig. \ref{fig:initialCondition}(c)] the latter is the most common scenario. An example for such an interaction is given in Fig. \ref{fig:bouncingStates}. Here we display a single boson in the wave packet state $|\text{G}\rangle = (1/\sqrt{2\pi\sigma^2})\sum_j e^{-i q_0 j}e^{(j-j_0)^2/2\sigma^2}|j\rangle$, where $j_0$, $q_0$, $\sigma$ are the initial central position, quasi-momentum, and width of the wave packet, respectively, impinging an immobile type I complex with $\Rc=2$. In much the same way that the dissipation acts to bind the bosons, it results in a hard core exclusion interaction between isolated bosons and complexes that in this example extends over $\Rc$ sites. In the case at hand this leads to an elastic collision with the type I complex essentially acting as a hard boundary. Using this mechanism one could imagine a situation where two immobile complexes enclose a boson, thereby acting as a trap.

More generally the range of the exclusion interaction is dependent on the internal state of interacting complexes. For the type I complex of two bosons with $\Rc = 3$ we define an effective complex-complex interaction as $H^{(\text{I})}_\text{int} = \lim_{W\to\infty} W \sum_{m>k,\lbrace\alpha,\beta\rbrace=1,2} \Theta(R+\alpha-|k-m|) n^{(\alpha)}_k\,n^{(\beta)}_m$, with $n^\alpha_k = b_k^{(\alpha)\dagger}b_k^{(\alpha)}$ and $\Theta(x)$ is the Heaviside step function (see Fig. \ref{fig:bouncingStates} for an illustration).

\textit{Outlook.---}
%We have shown that in a system of hard-core bosons the interplay between distance-selective particle loss and coherent hopping results in rich out-of-equilibrium dynamics. The quasi-stationary Zeno subspace reached from an initial Mott insulator state features two families of coherently bound complexes, that exhibit a number of interesting properties, such as effective SO coupling, flat dispersion relations and state-dependent interactions. 
In the future it will be interesting to study the quantum phases that emerge in systems that contain solely a single kind of complex, e.g. ones that feature state-dependent interactions and flat bands, and look at the case of a fermion system with equivalent dissipation. Such pure systems could be experimentally prepared in the ultra cold atoms lattice experiments discussed in Refs. \cite{Bakr2009,Fukuhara2013}.

\textit{Acknowledgements} We gratefully acknowledge insightful discussions with J.P. Garrahan regarding the initial fast relaxation dynamics and also Matteo Marcuzzi, Sam Genway and James Hickey regarding the dispersion relations. We furthermore thank Robin Stevenson for critical reading and comments on the manuscript. The research leading to these results has received funding from the European Research Council under the European Union's Seventh Framework Programme (FP/2007-2013) / ERC Grant Agreement No. 335266 (ESCQUMA). We also acknowledge financial support from EPSRC Grant no. \ EP/I017828/1.

\section{Appendices}
\subsection{Derivation of the effective master equation}
The effective master equation models the dynamics on Zeno subspace. We derive this effective master equation using the Kato resolvent method \cite{Garcia-Ripoll2009,Kato}. The form of our particular $\mathcal{L}_d$ allows us to decompose it into a set of eigenvalues, $k_i$, and eigenspaces or pseudo-projectors, $P_i$,
\begin{align}
\mathcal{L}_d = & \sum_i k_iP_i. \label{eqn:katoResolvant}
\end{align}
These projectors form a complete orthogonal set,
\begin{align}
P_iP_j = & \delta_{i,j}P_i, \label{eqn:projectorOrthogonality} \\
\sum_i P_i = & 1. \label{eqn:projectorSum}
\end{align}
The Zeno subspace has a corresponding zero eigenvalue, removing it from the expansion. Subbing Eq. \eqref{eqn:katoResolvant} into the master equation we get
\begin{align}
\dot{\rho} = & \mathcal{L}_c\rho + \sum_{\lambda} k_\lambda\rho_\lambda , \label{eqn:subbedMasterEquation}
\end{align}
where $\rho_i=P_i\rho$ and $\lambda$ omits the steady state space. As the steady state space is the one of interest we define the projector onto the irrelevant space as $Q=\sum_\lambda P_\lambda$. We split Eq. \eqref{eqn:subbedMasterEquation} into the evolution of the relevant and irrelevant spaces by applying the respective projectors:
\begin{align}
\dot{\rho_0} = & P_0 \mathcal{L}_c \rho_0 +P_0\mathcal{L}_cQ\rho, \label{eqn:rho0MasterEquation} \\
Q\dot{\rho} = & Q\mathcal{L}Q\rho + Q\mathcal{L}\rho_0,  \label{eqn:QMasterEquation} 
\end{align}
where $\mathcal{L}= \mathcal{L}_d+\mathcal{L}_c$. Formal integration of  Eq. \eqref{eqn:QMasterEquation} gives
\begin{align}
Q\rho(t) = e^{tQ\mathcal{L}}Q\rho(0) + \int_0^t d\tau e^{(t-\tau)Q\mathcal{L}}Q\mathcal{L}\rho_0(\tau). \label{eqn:QSolution}
\end{align}
We assume that we start entirely in the steady state space i.e. $Q\rho=0$. Expanding $L$ we show that Eq. \eqref{eqn:QSolution} becomes
\begin{align}
Q\rho(t) = \int_0^t d\tau e^{(t-\tau)Q\mathcal{L}}Q\mathcal{L}_c\rho_0(\tau).
\end{align}
Which is substituted into Eq. \eqref{eqn:rho0MasterEquation} to give
\begin{align}
\dot{\rho_0} = & P_0 \mathcal{L}_c \rho_0 +P_0\mathcal{L}_c\int_0^t d\tau e^{(t-\tau)Q\mathcal{L}}Q\mathcal{L}_c\rho_0(\tau).
\end{align}
Taking a Laplace transform of this equation gives
\begin{align}
\mathbb{L}[\dot{\rho_0}] = P_0 \mathcal{L}_c \mathbb{L}[\rho_0] + P_0\mathcal{L}_c \frac{1}{s - Q\mathcal{L}}Q\mathcal{L}_c\mathbb{L}[\rho_0].
\end{align}
We then use the fact that $\gamma\gg J$, implying that the amplitudes of the Liouvillians compare as $\mathcal{L}_d\gg\mathcal{L}_c$. This allows an expansion of $(s-Q\mathcal{L})^{-1}$ to second order:
\begin{align}
\mathbb{L}[\dot{\rho_0}] \approx P_0 \mathcal{L}_c \mathbb{L}[\rho_0] + P_0\mathcal{L}_c \frac{1}{s - Q\mathcal{L}_d}Q\mathcal{L}_c\mathbb{L}[\rho_0].
\end{align}
We then perform an inverse Laplace transform to give
\begin{align}
\dot{\rho_0}  \approx P_0 \mathcal{L}_c \rho_0(t) + P_0\mathcal{L}_c \int_0^t d\tau e^{(t-\tau)\mathcal{L}_d}Q\mathcal{L}_c\rho_0(\tau).
\end{align}
Expanding $\mathcal{L}_d$ in terms of its projectors and expanding the exponential, we find
\begin{align}
\dot{\rho_0}  \approx P_0 \mathcal{L}_c \rho_0(t) + \sum_\lambda P_0\mathcal{L}_c \int_0^t d\tau e^{(t-\tau)k_\lambda}P_\lambda\mathcal{L}_c\rho_0(\tau).
\end{align}
By integration by parts, this remaining integral is re-expressed as
\begin{align}
\dot{\rho_0}  \approx& P_0 \mathcal{L}_c \rho_0(t) + \sum_\lambda P_0\mathcal{L}_c [(\frac{-1}{k_\lambda}P_\lambda\mathcal{L}_c(\rho_0(t) + \rho_0(0)e^{tk_\lambda})) \notag\\ -& \frac{e^{tk_\lambda}}{k_\lambda}\int_0^t d\tau e^{-\tau k_\lambda}P_\lambda\mathcal{L}_c\frac{d\rho_0(\tau)}{d\tau}].
\end{align}
Due to $\mathcal{L}_d$ is a purely dissipative Liouvillian, the $k_\lambda$'s are all negative. As we are interested in the long time limit, $t\gg 1/\gamma$, the second term is considered negligible, as is the remaining integral due to it is of higher order in $J/\gamma$ as $\frac{d\rho_0(\tau)}{d\tau} \propto J$. Leaving an effective master equation with the form
\begin{align}
\dot{\rho_0} \approx& P_0 \mathcal{L}_c \rho_0(t) - \sum_\lambda \frac{1}{k_\lambda} P_0\mathcal{L}_c P_\lambda\mathcal{L}_c\rho_0(t). \label{eqn:effectiveMasterEquation}
\end{align}

\subsection{Derivation of the Projected Hamiltonian and Jump Operators}
Due to the form of Eq. \eqref{eqn:effectiveMasterEquation}, we are only interested in states which are coupled to the Zeno subspace via a single tunnelling event. This leads us to only study the cases of a single pair and a double pair, which shares the central boson, at the critical distance $\Rc$. We define the forms of the pseudo-projectors, $P_i$, of $\mathcal{L}_d$ on this truncated space as:
\begin{align}
P_0\rho = & Q_0 \rho Q_0 + \sum_i \sigma_i ^- \sigma_{i+\Rc}^- Q_1 \rho Q_1 \sigma_{i+\Rc}^+ \sigma_{i}^+ \notag\\
& + \sum_i \sigma_{i-\Rc}^- \sigma_i ^- \sigma_{i+x}^- Q_2 \rho Q_2 \sigma_{i+\Rc}^+ \sigma_{i}^+ \sigma_{i-\Rc}^+, \label{eqn:P0Projection} \\
P_1 \rho = & Q_0 \rho Q_1 + Q_1 \rho Q_0 , \label{eqn:P1Projection} \\
P_2 \rho = & Q_0 \rho Q_2 + Q_2 \rho Q_0  \label{eqn:P2Projection}
\end{align}
where: 
\begin{align}
Q_1 = & \sum_m n_m n_{m+\Rc}\prod_{i\neq m} (1-n_i n_{i+\Rc}), \label{eqn:Q1}\\
Q_2 = & \sum_m n_{m-\Rc}n_m n_{m+\Rc}\prod_{i\neq m,m-\Rc} (1-n_i n_{i+\Rc}). \label{eqn:Q2}
\end{align}
$Q_0$ was introduced previously and projects onto no pairs, $Q_1$ projects onto a single pair and $Q_2$ projects onto two pairs which share the central boson. The first projector $P_0$ is the steady state space of $\mathcal{L}_d$, $P_0 = \lim_{t\rightarrow \infty} \mathcal{L}_d$, it includes only states with no pairs of bosons at the critical distance. The next two higher order projectors, $P_1$ and $P_2$ include states with a single pair and a double pair which share a central boson. It can be checked that $P_0$, $P_1$ and $P_2$ project onto the eigenspaces of $\mathcal{L}_d$ with eigenvalues 0, $-\gamma/2$ and $-\gamma$ respectively.

The exact derivation of the projected Hamiltonian from the first term of \eqref{eqn:effectiveMasterEquation} relies on the assumption that the system starts in the steady state space, meaning that we reduce $\rho_0 = P_0\rho = Q_0\rho Q_0$, and the property of the $Q$'s, $Q_i Q_j = \delta_{i,j}Q_i$, allowing it to be found by the following method
\begin{align}
P_0 L_c P_0 \rho_0 = & -iP_0[H,\rho_0] \notag \\
= &-iP_0(HQ_0\rho Q_0 - Q_0\rho Q_0 H) \notag\\
= &-i(Q_0HQ_0\rho Q_0 - Q_0\rho Q_0 H Q_0) \notag\\
= &-i[Q_0 HQ_0,\rho]
\end{align}
Giving the form of $H_\text{Z}$ as quoted.

We then formulate the projected jump operators from the second term of \eqref{eqn:effectiveMasterEquation}. We first rewrite this term as:
\begin{align}
- \sum_\lambda &\frac{1}{k_\lambda}P_0L_cP_\lambda L_cP_0\rho_0(t) =  P_0(-\frac{2}{\gamma}[H,Q_1 [H,\rho_0]Q_0] \notag\\
&- \frac{2}{\gamma}[H,Q_0[H,\rho_0]Q_1] -\frac{1}{\gamma}[H,Q_2 [H,\rho_0]Q_0] \notag\\
 &- \frac{1}{\gamma}[H,Q_0[H,\rho_0]Q_2])
\end{align}
Which splits into two equations corresponding to the $k_\lambda$ eigenvalues
\begin{align}
-\frac{2}{\gamma}&(Q_0HQ_1HQ_0\rho_0 + \rho_0 Q_0HQ_1HQ_0 \notag\\
&-2\sum_j \sigma_j^- \sigma_{j+\Rc}^- Q_1HQ_0 \rho_0 Q_0HQ_1 \sigma_{j+\Rc}^+ \sigma_j^+ )\label{eqn:Q1Lindblad}\\
-\frac{1}{\gamma}&(Q_0HQ_2HQ_0\rho_0 + \rho_0 Q_0HQ_2HQ_0 \notag\\
&-2\sum_j \sigma_{j-\Rc}^- \sigma_j^- \sigma_{j+\Rc}^- Q_2HQ_0 \rho_0 Q_0HQ_2 \sigma_{j+\Rc}^+ \sigma_j^+ \sigma_{j-\Rc}^+) \label{eqn:Q2Lindblad}
\end{align}
Upon expansion of $Q_0 H Q_1$ and $Q_0 H Q_2$ we find a Lindblad form with the jump operators shown.

\subsection{Derivation of dispersion relations}
To demonstrate how the dispersion relations are calculated the single example of a type I state with 2 bosons for $\Rc=4$ will be shown. As stated, each site has an associated set of internal states, $|j,\alpha\rangle$, where $\lbrace\alpha\rbrace = 1\rightarrow 3$. We define a state of the system as $|\psi(j)\rangle = [A(j)|1\rangle + B(j)|2\rangle + C(j)|3\rangle]|j\rangle$ and perform a Fourier transform on this state to give the external quasi-momentum states $|K\rangle = (1/N)\sum_j e^{ijq_K} [A(K)|1\rangle + B(K)|2\rangle + C(K)|3\rangle]|j\rangle$

We rewrite the projected Hamiltonian in this basis as
\begin{align}
H_\text{Z} = & J\sum_{j=1}^{N}[ |j,1\rangle\langle j,2| + |j+1,1\rangle\langle j,2| \notag\\&+ |j,2\rangle\langle j,3| + |j+1,2\rangle\langle j,3| +\text{H.c.}]. \label{eqn:23Hamiltonian}
\end{align}
Applying this to the $|K\rangle$ state it is shown that you are it reduces to an operator on the spin structure:
\begin{align}
H_\text{Z}|K\rangle = J \begin{pmatrix} 0 & 1+e^{-iq_K} & 0 \\ 1+e^{iq_K} & 0 & 1+e^{-iq_K} \\ 0 & 1+e^{iq_K} & 0 \end{pmatrix}|K\rangle
\end{align}
Solving for the eigenvalues and eigenvectors of this matrix yields the results shown for the dispersion relations of this complex.

\bibliography{dissipativeBinding}

\end{document}